\documentclass[10pt,aps,prd,reprint,twocolumn,bibnotes,nofootinbib,amsfonts,amssymb,amsmath,preprintnumbers,notitlepage]{revtex4-1}
\usepackage[utf8]{inputenc}
\usepackage[british]{babel}
\usepackage[Symbolsmallscale]{upgreek}
\usepackage{isomath}
\usepackage[protrusion=true,tracking=true,kerning=true,spacing=true,final,babel=true]{microtype}
\usepackage{physics}
\usepackage{xcolor}
\usepackage{graphicx}
\usepackage[final]{hyperref}

\begin{document}
\title{Response to Cusin~et~al's comment on \texttt{arXiv:1810.13435}}

\author{Alexander~C.~Jenkins}
\email{alexander.jenkins@kcl.ac.uk}
\affiliation{Theoretical Particle Physics and Cosmology Group, Physics Department, King's College London, University of London, Strand, London WC2R 2LS, United Kingdom}

\author{Mairi~Sakellariadou}
\email{mairi.sakellariadou@kcl.ac.uk}
\affiliation{Theoretical Particle Physics and Cosmology Group, Physics Department, King's College London, University of London, Strand, London WC2R 2LS, United Kingdom}

\author{Tania~Regimbau}
\email{tania.regimbau@lapp.in2p3.fr}
\affiliation{LAPP, Universit\'e Grenoble Alpes, USMB, CNRS/IN2P3, F-74000 Annecy, France}

\author{Eric~Slezak}
\email{eric.slezak@oca.eu}
\affiliation{Universit\'e C\^ote d'Azur, Observatoire de la C\^ote d'Azur, CNRS, Laboratoire Lagrange, CS 34229, 06304 Nice Cedex 4, France}

\author{Richard~O`Shaughnessy}
\email{richard.oshaughnessy@ligo.org}
\affiliation{Center for Computational Relativity and Gravitation, Rochester Institute of Technology, 85 Lomb Memorial Drive, Rochester, NY 14623, USA}

\author{Daniel~Wysocki}
\email{dw2081@rit.edu}
\affiliation{Center for Computational Relativity and Gravitation, Rochester Institute of Technology, 85 Lomb Memorial Drive, Rochester, NY 14623, USA}

\date{\today}

\begin{abstract}
    We offer a brief response to the criticisms put forward by Cusin~et~al in Ref.~\cite{Cusin:2018ump} about our work Refs.~\cite{Jenkins:2018kxc,Jenkins:2018uac}, emphasising that none of these criticisms are relevant to our main results.
\end{abstract}
\maketitle

%%%%%%%%%%%%%%%%%%%%%%%%%%%%%%%%%%%%%%%%%%%%%%%%%%%%%%%%%%%%%%%%%%%%%%%%%%%%%%%%%

All of the criticisms raised by Cusin~et~al in Ref.~\cite{Cusin:2018ump} are about the analytical approach that we introduced in Ref.~\cite{Jenkins:2018uac}.
Indeed, we stressed explicitly ourselves throughout Ref.~\cite{Jenkins:2018uac} that this analytical approach is inaccurate.
In particular, after introducing the analytical expression, we wrote:
    \begin{quotation}
        ``We emphasize that Eq.~(69) [the analytical expression] is only a simple approximation of the true angular spectrum of the anisotropies\ldots
        We also note that we have extended Eq.~(64) [the power-law approximation of the galaxy-galaxy two-point correlation function] beyond its realm of validity by assuming that it holds for all distances $d$\ldots
        We therefore turn to a more detailed and accurate approach in the following section, to address the deficiencies of this simple model.''
    \end{quotation}

The main results of Refs.~\cite{Jenkins:2018kxc,Jenkins:2018uac} are \emph{not} based on this analytical approximation; they come from a thorough and careful analysis of a large simulated galaxy catalogue, based on the Millennium simulation~\cite{Springel:2005nw,Blaizot:2003av,Lemson:2006ee,DeLucia:2006szx}.
The analytical approach was only ever intended as a simplistic first pass at the problem, before doing the full catalogue analysis.
While the analytical approach can provide valuable insights (particularly for rapid investigations of e.g. tens of thousands of different astrophysical models~\cite{Jenkins:2018kxc}), it should not be regarded as a confident prediction, whereas the catalogue results should be.
It is unambiguously clear that none of the criticisms in Ref.~\cite{Cusin:2018ump} are relevant to our catalogue approach.

Thus, it remains to understand the difference between the results of our catalogue approach and the results of Cusin~et~al in Ref.~\cite{Cusin:2018rsq}.
In order to clarify the difference between the approaches, we list the steps that must go into a first-principles calculation of the $C_\ell$'s of the astrophysical stochastic background, and described how each of these is achieved for us and for Cusin et al:
    \begin{enumerate}
        \item The cosmological perturbations must be evolved from some initial power spectrum at early times, giving the dark matter overdensity field down to redshift zero.
        \begin{description}
            \item[Jenkins~et~al] This is done for us by the Millennium simulation, where the N-body gravitational clustering dynamics is simulated numerically, in a way that automatically accounts for all non-linear effects.
            \item[Cusin~et~al] This is done using the Boltzmann code \textsc{CMBQuick}, which treats the overdensities as small, linearised perturbations around a homogeneous background, and solves the linear evolution equations for these overdensities.
            The linear approximation is fine for CMB calculations at $z>10^3$ (which is what these types of codes were originally designed for), but breaks down at $z\approx0$, particularly on small scales.
            Cusin~et~al attempt to account for non-linear effects at $z\approx0$ by using the HALOFIT algorithm~\cite{Smith:2002dz}; this is essentially an ad-hoc fitting function, calibrated to N-body simulations.
            However, HALOFIT is known to under-estimate the matter power spectrum in $\Lambda$CDM.\footnote{From John Peacock's website where the code is hosted, \url{www.roe.ac.uk/~jap/haloes/} --- ``HALOFIT has received a lot of use, and has been incorporated into CMB packages such as CMBFAST and CAMB. Nevertheless, it is not perfect: it reflected accurately the state of the art of simulations as of 2003, but subsequent work has pushed measurements to smaller scales and higher degrees of nonlinearity. This has revealed that HALOFIT tends to underpredict the power on the smallest scales in standard LCDM universes (although HALOFIT was designed to work for a much wider range of power spectra).''}
            What's more, the simulations that HALOFIT is based on are older than Millennium, and smaller by a factor of $(2160/256)^3\approx600$.
        \end{description}
        \item The dark matter haloes must be populated with galaxies, accounting for the fact that the galaxies are more tightly clustered than the haloes themselves.
        \begin{description}
            \item[Jenkins~et~al] This is done for us by the simulated galaxy catalogue, which is based on a sophisticated semi-analytical model~\cite{DeLucia:2006szx} that accounts for a whole host of messy astrophysical feedback processes to model the galaxy population and distribution within each dark matter halo.
            \item[Cusin~et~al] This is done by writing the galaxy power spectrum as a biased form of the matter power spectrum, $P_\mathrm{gal}=b^2P_\mathrm{mat}$.
            The bias function $b$ is itself based on a linearised approximation, and is assumed to be scale-invariant (which is known to be false, particularly on small scales).
        \end{description}
        \item The gravitational-wave (GW) emission of each galaxy must be calculated, assuming a particular model for the compact binary populations.
        \begin{description}
            \item[Jenkins~et~al] This is done using detailed information from the Millennium simulation about the star formation rate in each galaxy as a function of time, which allows us to calculate the compact binary merger rate at the time of GW emission.
            We also use information about the metallicity and peculiar velocity of each galaxy, as these influence the observed GW flux.
            \item[Cusin~et~al] This is done using analytical formulae, treating the emitted GW flux of each galaxy as a function of the mass of the host halo only.
        \end{description}
    \end{enumerate}
Following steps 1--3, one can then superimpose the GW flux from all the galaxies to calculate the $C_\ell$'s.

\begin{figure*}[p!]
    \includegraphics[width=\textwidth]{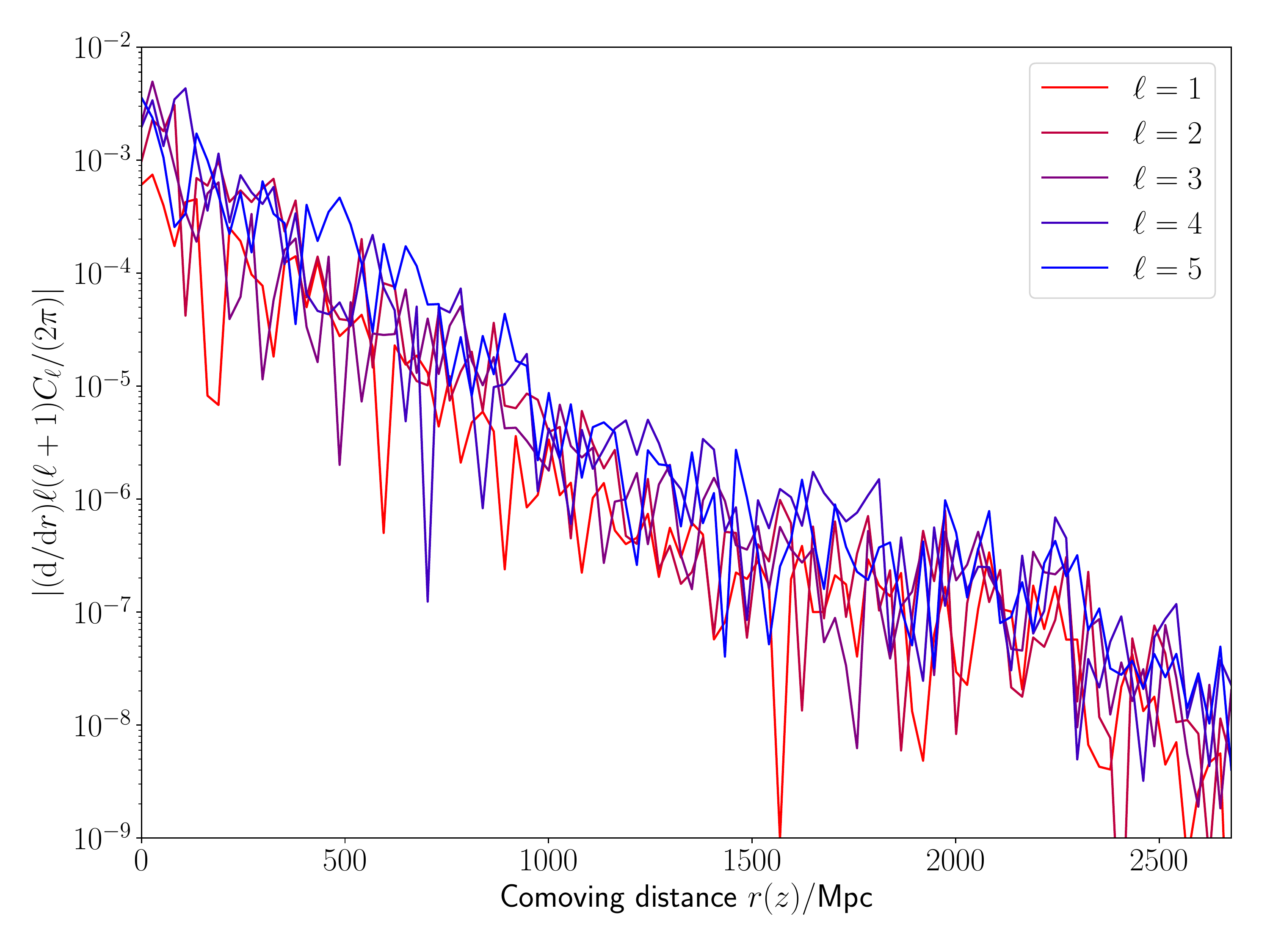}
    \caption{This shows the absolute change in $C_\ell$ as the maximum comoving distance of the galaxies is increased, effectively giving the relative contribution of each distance bin. The exponential decrease here illustrates that the anisotropies are dominated by the nearest galaxies, and are therefore sensitive to relatively small scales. We focus on the lowest $\ell$-modes, as these will be the easiest to detect with ground-based GW interferometers.}
    \label{fig:C_ell_diff}
\end{figure*}

\begin{figure*}[p!]
    \includegraphics[width=\textwidth]{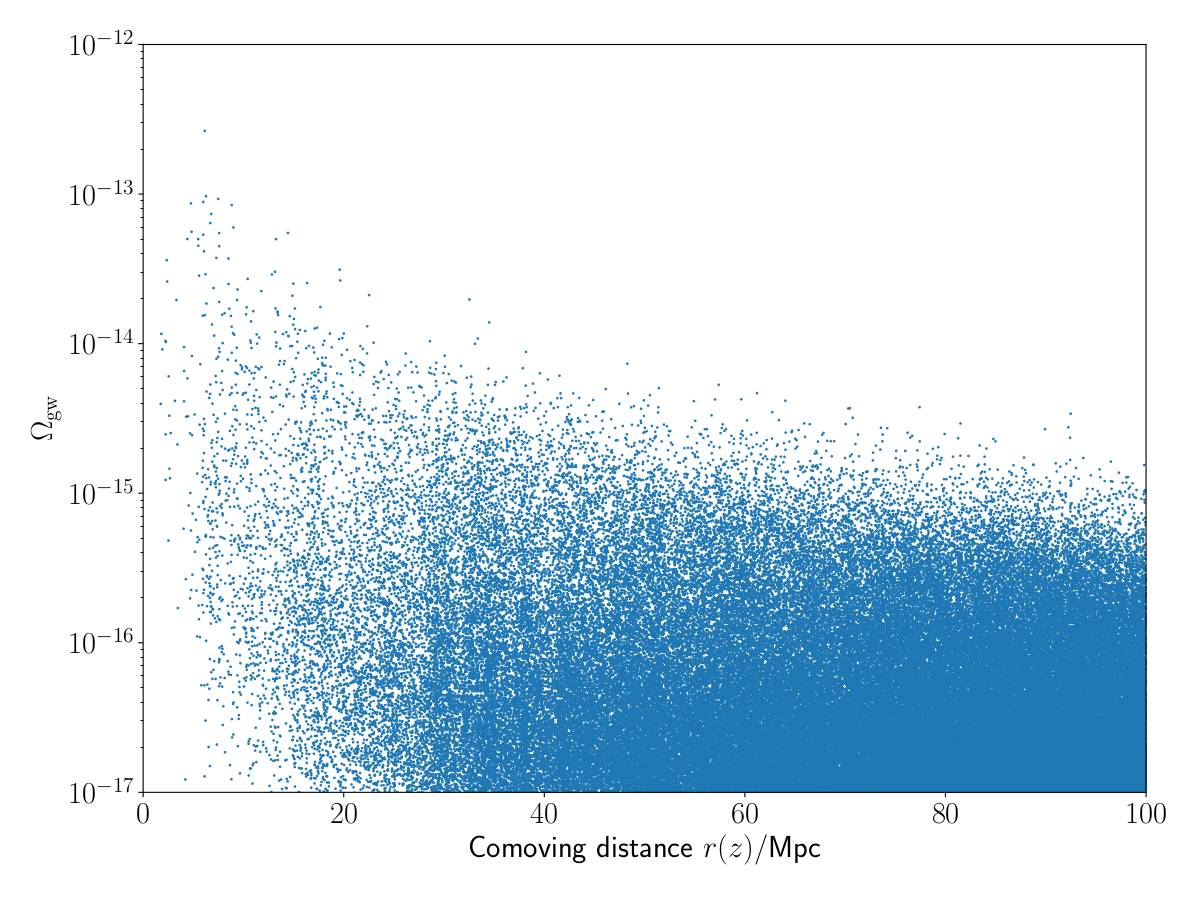}
    \caption{Each point represents one galaxy, showing its comoving distance in megaparsecs, and its contribution to $\Omega_\mathrm{gw}$. The GW-brightest galaxies are all within 10~Mpc.}
    \label{fig:brightest-galaxies}
\end{figure*}

In Ref.~\cite{Jenkins:2018kxc} we sample several thousand possible binary black hole (BBH) populations supported by the LIGO/Virgo O1 detections, and find that these only affect the $C_\ell$'s at a level of $\approx3\%$.
So it seems that step 3 is not the cause of the discrepancy between us and Cusin~et~al (they state that they agree with this in Ref.~\cite{Cusin:2018rsq}).
One must therefore look at the different approaches to steps 1 and 2.
For both steps, we have in the discussion above defended the accuracy of our catalogue approach relative to that of Cusin et al.
In particular, their use of linear perturbation theory and a linear, scale-invariant galaxy bias can be expected to lead to a loss of clustering on small scales.
This is important because, although we are interested in large angular scales (small $\ell$), most of the astrophysical background comes from low redshifts ($z\lesssim1$), and the anisotropies are dominated by the very lowest redshifts (this is because changing the number of galaxies in a given direction at low redshifts causes a much more significant fluctuation than at higher redshifts, as the GW flux from each individual galaxy is much greater---see Fig.~\ref{fig:C_ell_diff}).
All of the GW-brightest sources included in our catalogue are at distances less than 10~Mpc, where non-linear effects will be important (see Fig.~\ref{fig:brightest-galaxies}).
So even on large angular scales, the $C_\ell$'s are sensitive to GW sources at small distances from us, and therefore to small scales in the galaxy power spectrum.

To confirm that the linearised approximation (augmented with HALOFIT) adopted by Cusin~et~al in Ref.~\cite{Cusin:2018rsq} lead to smaller anisotropies, we compared in  Ref.\cite{Jenkins:2018kxc} the $C_\ell$'s obtained using the Millenium catalogue and that of Ref.~\cite{Cusin:2018rsq} for the same (maximum-likelihood) BBH distribution and fixing all other details of the astrophysical model to be the same.
We refer the reader to Fig.~2 of our paper~\cite{Jenkins:2018kxc}

In summary, we have pointed out that none of Cusin~et~al's criticisms in Ref.~\cite{Cusin:2018rsq} are at all relevant to our catalogue approach, which is the basis for all of our main results.
We have defended the accuracy of this approach compared to the linearised approach used by Cusin~et~al in Ref.~\cite{Cusin:2018rsq}.

%%%%%%%%%%%%%%%%%%%%%%%%%%%%%%%%%%%%%%%%%%%%%%%%%%%%%%%%%%%%%%%%%%%%%%%%%%%%%%%%%
\bibliography{response}
\end{document}